\def\gtsim{\hbox{\raise 2pt \hbox {$>$} \kern-1.1em \lower 4pt \hbox {$\sim$}}}

\documentclass{iaus}
\usepackage{graphicx}

\title[] 
{Relativistic plasma and \\ ICM/radio source interaction}

\author[L. Feretti et al.]   
{Luigina Feretti$^1$,
Gabriele Giovannini$^{1,2}$,
Federica Govoni$^{3}$,
\and Matteo Murgia$^{3}$}

\affiliation{$^1$INAF Istituto di Radioastronomia, \\ Via P. Gobetti n. 101, 
40129 Bologna, Italy \\ email: {\tt lferetti@ira.inaf.it} \\[\affilskip]
$^2$Dipartimento di Astronomia, Universit\'a di Bologna, \\ via Ranzani n.1, 
40127  Bologna, Italy \\[\affilskip]
$^3$INAF Osservatorio Astronomico di Cagliari, \\ Strada 54, Loc. Poggio 
dei Pini, 09012 Capoterra, Italy }

\pubyear{2011}
\volume{274}  
\pagerange{1--6}
\jname{Advances in Plasma Astrophysics}
\editors{A. Bonanno, E. de Gouveia Dal Pino \&  A. Kosovichev, eds.}
\begin{document}

\maketitle

\begin{abstract}
The first detection of a diffuse radio source in a cluster of
galaxies, dates back to the 1959 (Coma Cluster, \cite{lar59}).
Since then, synchrotron radiating radio sources have been found in
several clusters, and represent an important cluster component which is linked
to the thermal gas.  Such sources indicate the existence of large
scale magnetic fields and of a population of relativistic electrons in
the cluster volume.  The observational results provide evidence that
these phenomena are related to turbulence and shock-structures
in the intergalactic medium, thus playing a major role in the
evolution of the large scale structure in the Universe. The interaction
between radio sources and cluster gas is well established in particular
at the center of cooling core clusters, where feedback from AGN is a 
necessary ingredient to adequately describe the formation and evolution 
of galaxies and host clusters.
\end{abstract}

\firstsection 
\section{Introduction}

Clusters of galaxies are the largest gravitationally bound systems in
the Universe.  Most of the gravitating matter in any cluster is in the
form of dark matter ($\sim$ 80\%). Some of the luminous matter is in
galaxies ($\sim$ 3-5\%), the rest is in diffuse hot gas ($\sim$
15-17\%), detected in X-ray through its thermal bremsstrahlung
emission.  
This thermal plasma, consisting of particles of energies of several
keV, is commonly referred to as Intracluster Medium (ICM).  In recent
years it has become clear that the ICM can also contain highly
relativistic particles, whose number density is of the order of
10$^{-10}$ cm$^{-3}$. Although the relativistic plasma has an energy
density of $<$1\% than that of the thermal gas, it is nevertheless
very important in the cluster formation and evolution.

Clusters are formed by hierarchical structure formation processes.  In
this scenario, smaller units formed first and merged to larger and
larger units in the course of time.  The merger activity appears to be
continuing at the present time, and explains the relative abundance of
substructure and temperature gradients detected in Abell clusters by
optical and X-ray observations.  At the end of their evolution,
clusters reach a relaxed state, with a giant galaxy at the center, and
enhanced X-ray surface brightness peak in the cores.  The hot gas in
the centre has a radiative cooling time shorter than the expected
cluster age, therefore energy losses due to X-ray emission are important
and lead to a temperature drop towards the centre (\cite{fab94}). The
relaxed clusters are thus referred to as cooling core clusters.

\begin{figure}[b]
\begin{center}
 \includegraphics[width=5.4in,angle=270]{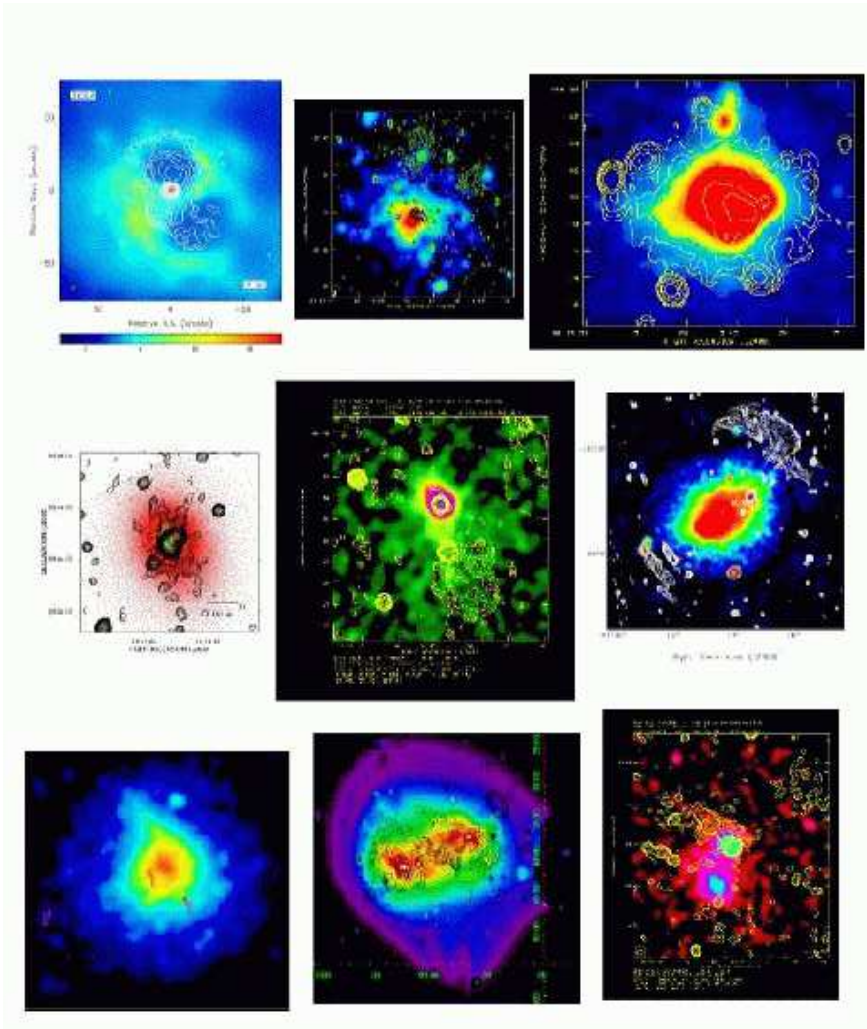} 
 \caption{Collection of clusters showing several types of 
radio emission, shown in contours, overlaid onto 
the X-ray emission, shown in colors. Clusters are (from left to right and
from top to bottom) Perseus (mini-halo and X-ray cavity), A\,548b (relic), 
A\,2163 (halo), A\,2029 (mini-halo), A\,1664 (relic), A\,3667 (double relics),
A\,119 (radio galaxies), A\,754 (halo plus relic), A\,115 (relic).}
   \label{fig1}
\end{center}
\end{figure}

From the radio point of view, clusters can host diffuse radio
emission, which has been now revealed in several conditions (merging
and relaxed clusters), at different cluster locations (center,
periphery, intermediate distance), and on very different size scales
(100 kpc to $>$Mpc),(see Fig.\,\ref{fig1} for several examples).
All diffuse radio sources have in common the short lifetimes
of the radiating particles, which therefore need to be
reaccelerated.
The properties of the radio emission are linked
to those of the host cluster, therefore the connection between the
thermal and relativstic plasma in clusters of galaxies is important
for the cluster formation and evolution.  The understanding of magnetic
field and relativistic particle properties
is important for a comprehensive physical
description of the intracluster medium in galaxy clusters.

\section{Diffuse cluster radio sources in merging clusters}

\begin{figure}[b]
\begin{center}
 \includegraphics[width=4.8in]{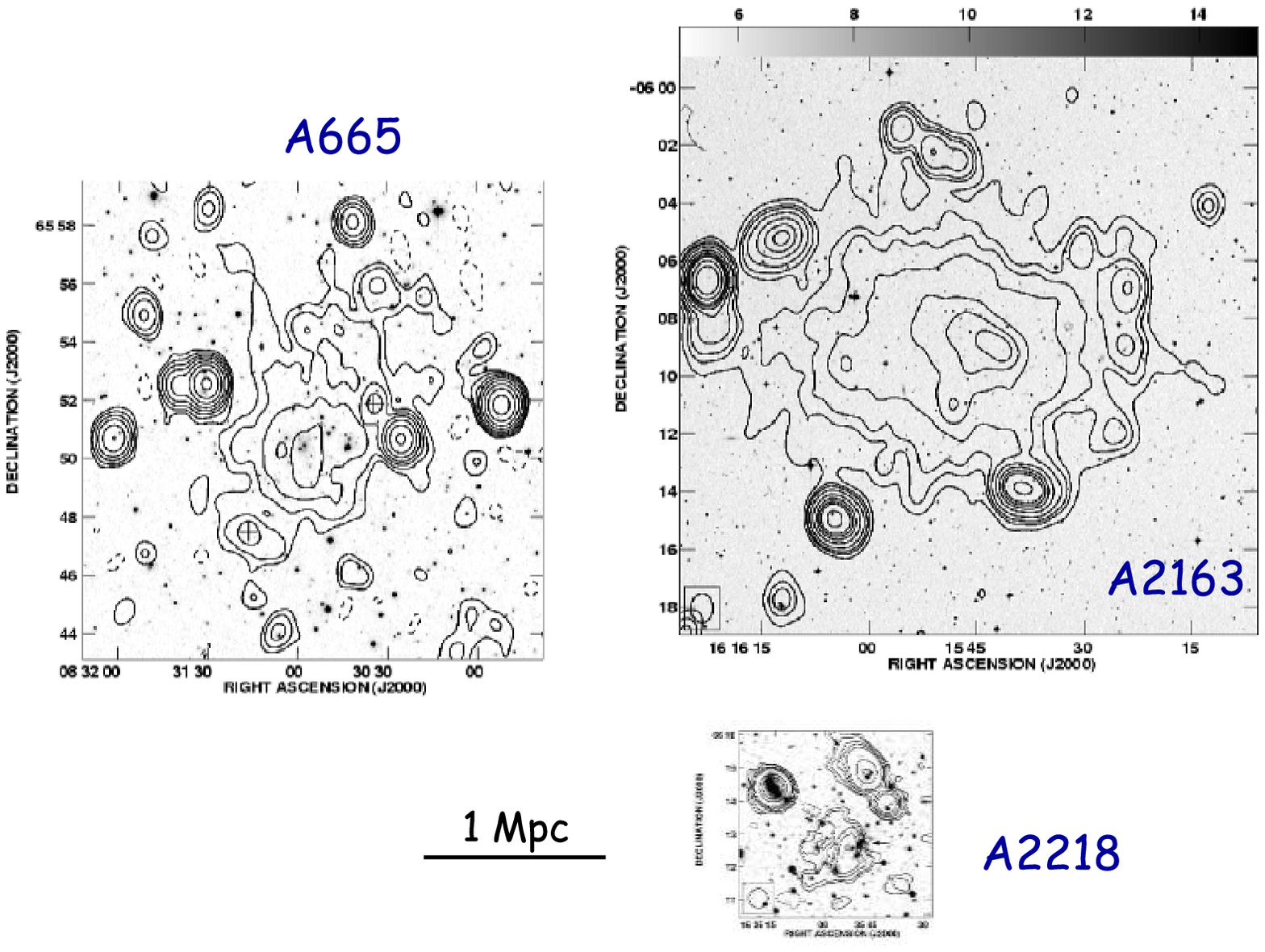} 
 \caption{Images of the 
clusters A\,665, A\,2163 and A\,2218, hosting radio halos: 
radio emission is represented by contours, which are overlaid onto 
the optical image. The maps are all scaled to the same linear scale.}
   \label{fig2}
\end{center}
\end{figure}

The most spectacular example of diffuse cluster radio sources is
represented by giant radio halos. They are associated with 
clusters undergoing merging processes, and believed to be
energized by  the turbulence produced
during the cluster mergers (see (\cite{fer08} for a review).
Radio halos (Fig.\,\ref{fig2}) can reach a size of 1 - 2  Mpc and more, 
although smaller size halos have also been found.
New halos have recently been detected in A\,851, 
A\,1213, A\,1351, A\,1995, A\,2034 and
A\,2294 (\cite{gio09}, also \cite{gia09} for A\,1351).
The most powerful radio halo known so far is found in the 
distant cluster MACS\,J0717.5 +3745 (\cite{bonb09}, \cite{wee09}) 
at z = 0.55.
A peculiar example of a double radio halo in a close pair of
galaxy clusters is represented by A\,399 and A\,401 (\cite{mur10}),
where the two radio halos could be either originated by previous merger 
histories of the two clusters, or due to the currently ongoing interaction.

The radio power of both small and giant halos correlates with the
cluster X-ray luminosity, i.e. gas temperature and total mass
(\cite{cas06}, \cite{gio09}), in the sense that 
highly luminous X-ray clusters host the most powerful radio halos.
The radio halos are generally associated with clusters with
X-ray luminosity in the 0.1-2.4 keV range $>>$ 10$^{44}$ erg s$^{-1}$.
Recently, however, radio halos have been found also in clusters
with X-ray luminosity around 10$^{43}$ erg s$^{-1}$,  which 
are typical of low density environments: 
the first example is the cluster A\,1213 (\cite{gio09}), other cases
are presented by \cite{bro09}.

\begin{figure}[b]
\begin{center}
 \includegraphics[width=3.4in]{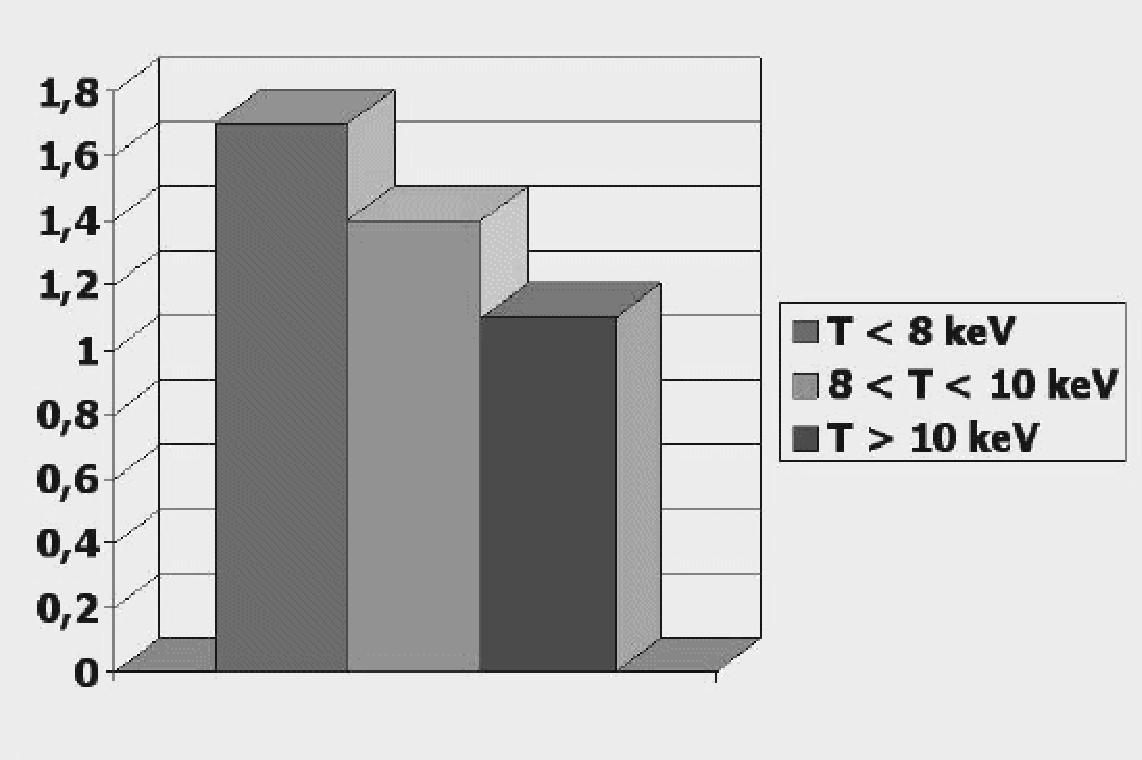} 
 \caption{Average spectral index of radio halos for clusters in
different ranges of
temperature, showing that hotter clusters tend to host halos with flatter
spectra.}
   \label{fig3}
\end{center}
\end{figure}

Another important link between the relativistic and thermal plasma is
represented by the connection between the cluster temperature and
the radio halo spectral index, first suggested by \cite{fer04}, 
and now confirmed  by \cite{gio09}. It is found that 
clusters at higher temperature tend to host halos with flatter
spectra (see Fig.\,\ref{fig3}). This correlation reinforces
the connection between radio emission and
cluster mergers, since hot clusters are more massive and may derive
from more energetic merging processes, supplying more energy to the
radiating electrons.

Other diffuse radio sources associated with cluster mergers 
are relics, located in cluster peripheral regions,
and characterized by high polarized emission
(see \cite{gio04} for a review and references).
Relics can be also detected in clusters containing radio halos. 
A very narrow giant relic, which displays highly aligned magnetic field,
has been recently detected by \cite{wee10}.
Remarkable are the giant double relics,  located on opposite sides 
with respect  to the cluster center,
whose prototype is A\,3667 (\cite{rot97}).
Several other cases have been found more recently: RXCJ1314.4-2515
(\cite{fer05}), A\,3376
(\cite{bag06}), A\,1240 and A\,2345 (\cite{bona09}).
Radio relics are likely energized by shock waves occurring during the
cluster mergers, as confirmed by observational
results obtained by e.g. \cite{sol08}, and \cite{fin10}.  

Diffuse radio emission is also detected beyond clusters, on very large
scales. An example is represented by the filament of galaxies ZwCl
2341.1+0000 at z $\sim$ 0.3, first detected by \cite{bag02}, and
fecently confirmed by \cite{gio10}, who also detect polarized emission.  
Noticeable are the complex radio structures detected on large
scale at more than 2 Mpc from the center of A\, 2255 (\cite{piz08}),
wich have been suggested to be connected with large scale structure
formation shocks.

\begin{figure}[b]
\begin{center}
 \includegraphics[width=5.4in]{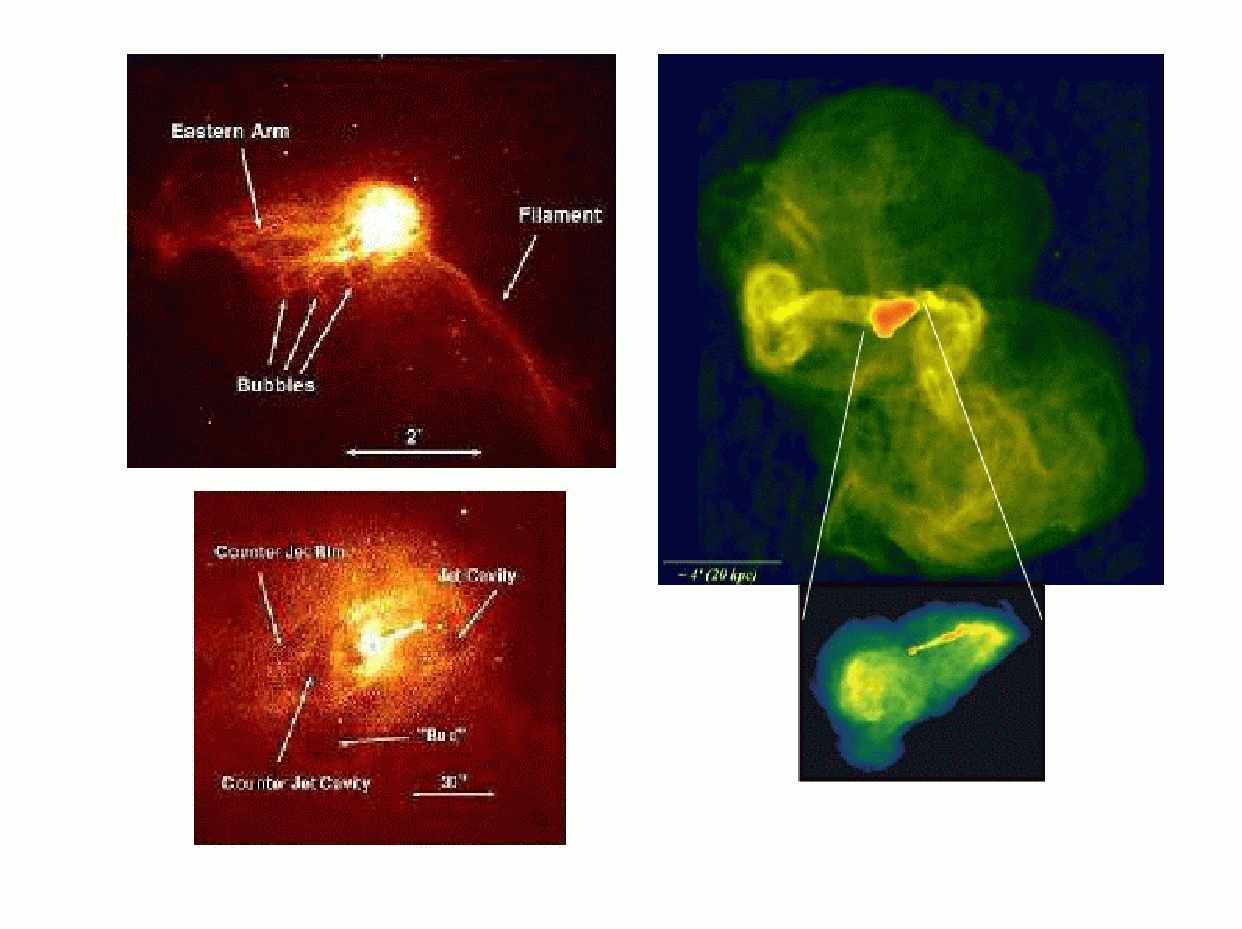} 
 \caption{X-ray and radio structures of M\,87, a buoyant bubble
in the center of a cool core cluster. The 2 left panels show
the X-ray filaments and cavities detected with Chandra (\cite{for07}).
The right panels show the large scale radio bubbles, with the radio jets
in the inset (\cite{owe00}). The angular scale is indicated 
within each image. }
   \label{fig4}
\end{center}
\end{figure}

\section{Radio - X-Ray interaction in cooling core clusters}

The diffuse radio sources which may be detected at the center of 
cooling core clusters are classified as  mini-halos, since they 
are morphologically similar to giant halos associated with
merging clusters, but they are smaller in size (a few hundreds kpc).
Although these sources are generally surrounding a powerful
central radio galaxy, it has been argued
that the energetics necessary to their maintenance is not
supplied by the radio galaxy itself, but the electrons are
reaccelerated by MHD turbulence in the cooling core region
(\cite{git02}).
A peculiar case is represented by the cluster RXJ\,1347.5-1145,
one of the highest X-ray luminous clusters known so far.
It is a relaxed clusters hosting a mini-halo, but it is also
characterized by minor mergers in the cluster periphery.
The mini-halo, detected by \cite{git07}, shows an asymmetrical
structure, with elongation coincident with a X-ray subclump,
suggesting that additional energy for the electron re-acceleration 
might be provided by the sub-merger event. 

In some cooling core clusters, as A\,13, A\,85,
A\,133, A\,4038, diffuse radio sources offset from the cluster
center have been detected (\cite{sle01}).
They can be classified as mini relics, because of their 
small-intermediate size.
They are characterized by very steep spectra ($\alpha \gtsim$ 2),
strong polarization, and filamentary  structure 
when observed with sufficient resolution. The phenomenology of diffuse 
 mini halos and mini
relics, may be related to the radio source/ISM interaction in the
central regions of cooling core clusters.

A spectacular example of the interaction between radio sources and the
hot intracluster medium is represented by cavities in the X-ray gas
distribution, filled with radio plasma. The first case has been
detected by \cite{boh93} in the ROSAT image of NGC\,1275 (3C\,84) in
the Perseus cluster. Here the thermal plasma is displaced by the inner
parts of the radio lobes, causing a significant decrease of the X-ray
surface brightness in those regions.  The high spatial resolution of
the Chandra X-ray Observatory has allowed the detection of X-ray
cavities in the inner region of many cooling core clusters. These
systems have been extensively studied in recent years, and a clear
correspondence between regions of radio emission and deficits in the
X-ray has been found in several cases (see \cite{bla10} and references
therein).  The interaction between the AGN jets and the ICM is
believed to be the primary feedback mechanism between the AGN driven
by supermassive black holes and its environment.


Cooling of the hot intracluster medium in cluster centers can feed the
supermassive black holes found in the nuclei of the dominant cluster
galaxies, leading to AGN outbursts, which can reheat the gas. 
The relativistic radio jets of the associated radio sources 
are creating cavities, through one strong episode or several small
episodes of energy release.  AGN heating can come in the form of
shocks, and buoyantly rising bubbles that have been inflated by radio
lobes (\cite{bes07}, \cite{bir08}).

A well studied radio source is M\,87 (\cite{owe00}) in the nearby
Virgo cluster (Fig.\,\ref{fig4}), where \cite{for07} show X-ray
structure and filaments which are related to the radio emission.
Radio bubbles, blown into the cluster gas, rise buoyantly, expand and
give rise to a complex structures. Radio jets are confined within
cavities at early times, but at later times the cosmic rays
diffuse into the surrounding hot gas to form radio lobes with sizes
much larger than the cavities.  According to \cite{mat08}, the bubbles
may eventually expand to the cluster outskirts, where the cosmic rays
would impact into the surrounding medium, giving rise to mini relics,

\section{Magnetic fields in clusters}

The presence of diffuse radio sources in clusters demonstrates the
existence of magnetic fields in the intracluster medium.  An
indipendent way to obtain information on the cluster magnetic field is
from the analysis of Faraday Rotation. This kind of studies has
allowed a breakthrough in the knowledge of the strength and structure
of magnetic fields in clusters of galaxies, by analyzing the Rotation
Measure of sources seen through the magnetized cluster medium (see
e.g. \cite{gov04}).
Most recent studies are those on the Coma Cluster
(\cite{bona10}), A\,665 (\cite{vac10}) and the radio galaxy 3C\,449
(\cite{gui10}). The results obtained so far can be summarized as
follows (see also \cite{bonb10}): 
(i) magnetic fields are present in all clusters; (ii) at the
center of clusters undergoing merger activity the field strenght is
around 1 $\mu$G, whereas at the center of relaxed cooling core
clusters the intensity is much higher ($\sim$ 10 $\mu$G); (iii) a
model involving a single magnetic field coherence scale is not
suitable to describe the observational data, because of different
scales of field ordering and tangling.

Assuming a magnetic field power spectrum: $|B_{\kappa}|^2 \propto
{\kappa}^{-n}$ (\cite{mur04}), the range of spatial scales is
found between $30-500$ kpc and the spectral index $n$ is in the
range $2-4$. In A\,2255, \cite{gov06}, from the comparison between
the observations and the simulation of both the radio halo and the
polarization of radio galaxies, obtain a flatter power spectrum
at the center, and a steeper power spectrum  at the periphery.
This could originate from the different turbulence development in
central and peripheral cluster regions.

The cluster magnetic field intensity shows a radial decline linked to
the thermal gas density n$_e$ as $B \propto $n$^x_e$. A trend with $x$
= 1/2 is expected if the B field energy scales as the thermal energy,
while $x$ = 2/3 if the B field results from a frozen-in field during
the cluster collapse.  The values of $x$ derived so far are in the
range $0.5-1$, thus not conclusive.

\section{Summary and future prospects}

The presence of relativistic plasma in clusters of galaxies is
demonstrated by the existence of halos and relics detected in merging 
clusters, and the mini-halos and mini relics in cooling core clusters.
These features, which were detected in the past only in a few
clusters, are now becoming more and more common, in particular,
it is remarkable that diffuse structures, on size scales of
hundreds kpc are detected in low density environments.
Diffuse radio emission is also detected in filaments and in the
large scale structure.  
X-ray cavities filled by radio bubbles are found at the center
of cooling core clusters and represent the extreme example of
interaction between the thermal and relativistic plasma.

Future prospects for the study of new radio sources of the different
classes, in particular at low powers and at large redshifts, and study
of the polarization come from the new generation radio telescopes,
like the eVLA, LOFAR, the SKA precursors ASKAP and MeerKAT, and
finally SKA. The study of relics and large scale diffuse sources is
promising to trace the large scale structure formation and the cosmic
web. Observations in the X-ray and optical domain are required to
establish the cluster conditions, the merger evolutionary stage, the
presence and properties of shocks, and the signatures of cluster
turbulence.  The comparison between the observational results and the
theoretical expectations will be crucial to understand the physical
link between the relativstic and thermal plasma.

\begin{acknowledgements}

L.F. wishes to thank the conference organizers for the invitation 
to this very stimulating meeting.

\end{acknowledgements}

\end{document}